\documentclass[conference]{IEEEtran}
\IEEEoverridecommandlockouts

\usepackage{graphicx,epstopdf}
\usepackage{balance}

\usepackage{amsmath,amssymb,amsfonts,dsfont}
\usepackage{algorithmic}
\usepackage{algorithm}
\usepackage[caption=false]{subfig}

\begin{document}
%
\title{Federated Multi-Agent DRL for Radio Resource Management in Industrial 6G in-X subnetworks

\thanks{This work is partly supported by the HORIZON-JU-SNS-2022-STREAM-B-01-02 CENTRIC project (grant agreement No. 101096379) and the HORIZON-JU-SNS-2022-STREAM-B-01-03 6G-SHINE project (grant agreement No. 101095738).}
}

\author{Bjarke~Madsen and~Ramoni~Adeogun\\
\textit{Department of Electronic Systems, Aalborg University, Denmark} \\
E-mail:ra@es.aau.dk

}

%



\maketitle

\begin{abstract}
Recently, 6G in-X subnetworks have been proposed as low-power short-range radio cells to support localized extreme wireless connectivity inside entities such as industrial robots, vehicles, and the human body. Deployment of in-X subnetworks within these entities may result in rapid changes in interference levels and thus, varying link quality. This paper investigates distributed dynamic channel allocation to mitigate inter-subnetwork interference in dense in-factory deployments of 6G in-X subnetworks. This paper introduces two new techniques, Federated Multi-Agent Double Deep Q-Network (F-MADDQN) and Federated Multi-Agent Deep Proximal Policy Optimization (F-MADPPO), for channel allocation in 6G in-X subnetworks. These techniques are based on a client-to-server horizontal federated reinforcement learning framework. The methods require sharing only local model weights with a centralized gNB for federated aggregation thereby preserving local data privacy and security. Simulations were conducted using a practical indoor factory environment proposed by 5G-ACIA and 3GPP models for in-factory environments. The results showed that the proposed methods achieved slightly better performance than baseline schemes with significantly reduced signaling overhead compared to the baseline solutions. The schemes also showed better robustness and generalization ability to changes in deployment densities and propagation parameters.
\end{abstract}


\IEEEpeerreviewmaketitle
\vspace{-4pt}\section{Introduction}
Wireless communication has evolved significantly, with advancements in 5G technology enabling faster and more reliable communication. However, the need for more efficient and reliable networks has led to the development of 6G technology, which promises ultra-reliable communication, higher data rates, and lower latency. Wireless communication has limitless applications, including virtual and augmented reality in entertainment, education, healthcare, and manufacturing, personalized healthcare solutions, and real-time environmental monitoring. To achieve these benefits, new communication technologies must be developed, such as 6G short-range low-power in-X subnetworks \cite{alma9921021867205762}, which can support demanding requirements inside entities like robots, production modules, vehicles, and human-body scenarios. Ensuring data privacy and security is crucial for advanced wireless communication networks, especially for sensitive personal or environmental data. Techniques for resource allocation must intelligently guarantee privacy and security while adapting to changing wireless environments.

\noindent\textbf{Related work:} Interference management solutions for wireless systems have been proposed in the literature, see e.g., \cite{sun2018learning}. There has also been a significant amount of work targetting 6G in-X subnetworks within the last few years. These solutions can be categorized into heuristics, machine learning (ML) methods, and reinforcement learning (RL) techniques. Heuristic methods generate near-optimal solutions based on simple rules or algorithms, providing good results with low computational complexity. In \cite{litC1}, three heuristic-based algorithms are presented: $\epsilon$-greedy channel selection, nearest neighbor conflict avoidance, and minimum Signal to Interference plus Noise Ratio (SINR) guarantee. Centralized Graph Coloring (CGC) is presented as a benchmark algorithm, but it is impractical due to its high signaling and computation overhead. A distributed interference-aware dynamic channel selection is presented in \cite{H1_alma9921082042605762}. In \cite{litC3}, channel selection is approached based on centralized selective graph constructions. ML-based solutions may offer a higher quality of service compared to heuristics by leveraging historical data and adapting to evolving network conditions. 

In \cite{ML0_alma9921018979305762}, a Deep Neural Network (DNN) was successfully trained in offline simulations using CGC with mobile subnetworks. Subsequently, this DNN was deployed for real-time distributed channel selection. A novel solution for centralized power control is presented in \cite{ML1_alma9921343555905762}, where decisions are based on positioning information using Graph Neural Network (GNN). 

In \cite{2_MARL}, joint allocation of channel and transmit power based on a distributed multi-objective optimization problem is addressed, and an approach of Q-learning for multiple agents based on limited sensing information is proposed. The GA-Net framework presented in \cite{MARL3_DuXiao2023MRLf} is a distributed framework Multi-Agent Reinforcement Learning (MARL) resource management, based on GNN. 

\noindent\textbf{Contributions:}
This paper investigates the potential of federated MARL techniques for efficient, privacy-preserving channel allocation in 6G in-X subnetworks. The authors present their first study on federated MARL for radio resource management (RRM) in short-range low-power 6G in-X subnetworks, presenting the first contribution on FRL for RRM in 6G in-X subnetworks. They propose two novel federated MARL solutions - F-MADDQN and F-MADPPO - for dynamic channel allocation in 6G in-X subnetworks. These techniques enable privacy-preserving collaborative training of a Double Deep Q-Network (DDQN) policy and proximal Policy Optimization (PPO) agent for dynamic channel selection by multiple subnetworks in the presence of an umbrella network. The training phase requires only offline connectivity to the umbrella network. The paper also presents performance evaluation in industrial factory environment simulations based on a realistic factory floor plan from 5G-ACIA and 3GPP channel models for in-factory environments.


\vspace{-5pt}\section{System Model and Problem Formulation}\label{sec:2_system_model_and_problem_formulation}
\noindent\textbf{System Model:}
We consider downlink transmissions in a wireless network comprising $N$ independent and mobile subnetworks, each responsible for serving one or more devices, such as sensors and actuators. Each subnetwork contains an Access Point (AP) which is responsible for managing transmissions with its associated devices. We denote the collection of subnetworks as $\mathcal{N} = {1, \cdots, N}$, and the set of devices in the $n^\text{th}$ subnetwork as $\mathcal{M}_n = {1, \cdots, M_n}$. It is assumed that each subnetwork's AP is equipped with a local Resource Manager (RM), which utilizes data acquired from wireless environment sensing or received from its devices. 
Within each mobile subnetwork, wireless transmission occurs over one of $K$ shared orthogonal channels. Given the limited availability of resources, the number of bands is typically much less than the number of co-existing subnetworks. We assume that transmissions within each subnetwork are orthogonal, hence there is no intra-subnetwork interference. In practical systems, it may be impossible to make transmissions completely orthogonal due to limitations on the bandwidth of the channel allocated to the subnetwork leading to intra-subnetwork interference. Consideration of the effects of such interference is however left for future work.   

The received SINR, denoted $\gamma_{n,m}$, on a link between the $n^\text{th}$ AP and $m^\text{th}$ device can be expressed
\begin{equation}
    \gamma_{n,m}(t) = \dfrac{g_{n,m}(t)}{\sum_{i\in\mathcal{I}_k}g_{n,i}(t) + \sigma^2},\label{eq:sinr}
\end{equation}
where $g_{n,m}(t)$ is the received power on the link between the $n$th AP and device $m$, $\mathcal{I}_k$ denotes the APs or devices transmitting on channel $k$ and $\sigma^2$ denotes the noise power and is defined as
    $\sigma^2 = 10^{(-174 + \text{NF} + 10\log_{10}(B_k))/10}$,
where NF denotes the noise figure and $B_k$ is the bandwidth of the channel.
%
We consider practical 6G in-X subnetworks featuring short packets (in the order of tens of bytes) transmission, making the infinite block length assumption and the Shannon approximation used in most of the existing studies unrealistic. 
To compute the capacity, we instead use the finite block-length approximation as \cite{9580341}
\begin{equation*}
    r_{n,m} = B_k\left[
    \log_2\left(1 + \gamma_{n,m}(t)\right) - \sqrt{\frac{V_{n,m}(t)}{l}} Q^{-1}(\epsilon)\log e
    \right], \label{eq:PF_throughput}
\end{equation*}
where $Q$ is the complementary Gaussian cumulative distribution function based on the code-word decoding error probability $\epsilon$, $\log e\approx 0.434$ is a constant constraint of the loss, and $V$ is the channel dispersion defined as \cite{9580341}.
\begin{equation}
    V_{n,m}(t) = 1 - \dfrac{1}{\left(1 + \gamma_{n,m}(t)\right)^2}.
\end{equation}
When the length of the code-word block $l$ tends toward infinity, the achieved rate approaches the classical Shannon's approximation.

\noindent\textbf{Problem Formulation:}
We focus on a resource allocation problem that involves the distributed selection of channels to maximize the achieved sum rate while ensuring that devices within each subnetwork achieve a specified minimum rate. This optimization problem can be mathematically expressed as
\begin{equation}\label{eqPrb}
{\text{P}}:\left\{ {{{\max }_{\left\{ {{{\mathbf{c}}^t}} \right\}}}\sum\limits_{m = 1}^M {{r _{n,m}}} \left( {{{\mathbf{c}}^t}} \right)} \right\}_{n = 1}^N{\text{ st: }}\:{r _{n,m}} \geq {r_{{\text{target }}}},\:\forall n,m
\end{equation}
Here, $\mathcal{K} = \left[1, \cdots, K\right]; k \in \{1, 2, \cdots, K\}$ represents the vector of channel indices selected by all subnetworks at time t. $r_{\text{target}}$ denotes the target minimum rate, assumed to be the same for all subnetworks. The problem described in \eqref{eqPrb} involves the joint optimization of $N$ conflicting non-convex objective functions, making it a challenging problem to solve. In this paper, federated MARL methods are proposed as a solution to this problem.
\begin{figure*}[ht]
\centering
    \includegraphics[width=1.0\textwidth]{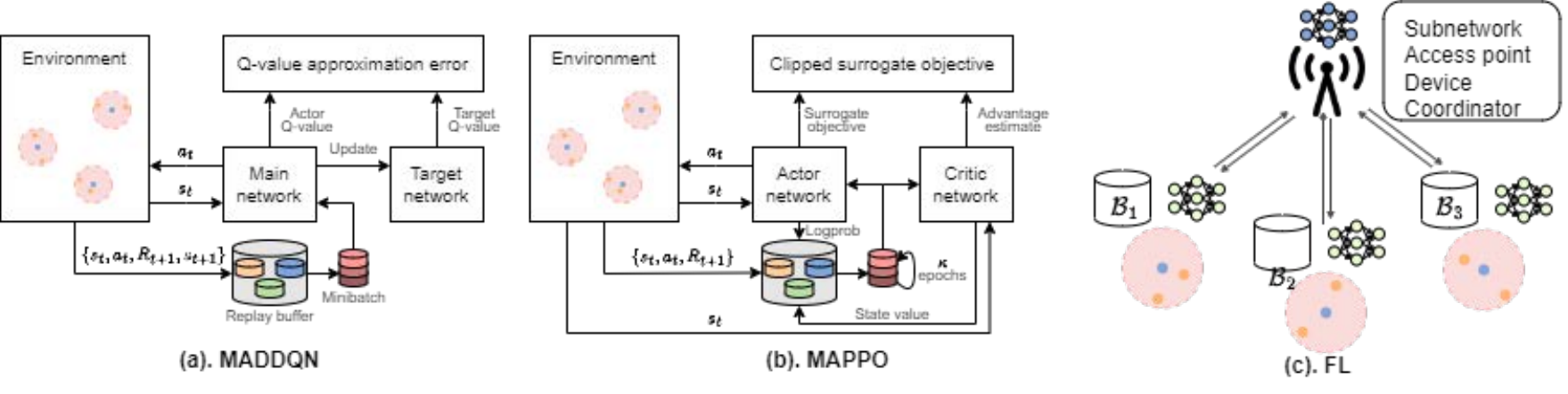}
    \caption{Illustration of the training procedures for (a) multi-agent DDQN and (b) multi-agent PPO and the federated learning concept (c).}
    \label{fig1}
\end{figure*}
\section{MARL FOR RRM IN SUBNETWORKS}\label{sec:3_MARL_FOR_SUBNETWORK_INTERFERENCE_MANAGEMENT}
\subsection{MARL Algorithms} \label{sec:marlalgorithms}
Resource selection in scenarios with multiple in-X subnetworks is cast as a decentralized partially observable Markov decision process (DEC-POMDP). A DEC-POMDP is formally represented as $\left(N,\mathcal{S},\{\mathcal{A}^n\}_{n=1}^N,\mathcal{P},\{\mathcal{R}^n\}_{n=1}^N\right)$.
The set of all possible states for all agents is defined as the state space $\mathcal{S}=\mathcal{S}_1\times\cdots\times\mathcal{S}_N$, and the joint action space containing all possible actions for the $n^\text{th}$ agent $\mathcal{A}^n$ is denoted $\pmb{\mathcal{A}}=\mathcal{A}^1\times\cdots\times\mathcal{A}^N$.
The reward signal for the $n^\text{th}$ agent is denoted $\mathcal{R}^n:\mathcal{S}\times\pmb{\mathcal{A}}\times\mathcal{S}$ and the transition probability from state $s\in\mathcal{S}$ to state $s'\in\mathcal{S}$ by joint action $\pmb{a}\in\pmb{\mathcal{A}}$ is denoted $\mathcal{P}=\mathcal{S}\times\pmb{\mathcal{A}}$.

\subsubsection{MADDQN}

Multi-Agent Double Deep Q Network (MADDQN) \cite{du2021survey, feriani2021single}  extends the Double Deep Q Network (DDQN) algorithm to enable decentralized learning among multiple agents. MADDQN allows each agent to have its own local Q-network and learn from its individual experiences. To train the Q-network, each agent interacts with the environment and collects experiences by executing actions according to its current policy. The experiences, including the state, action, reward, and next state, are stored in either a shared replay buffer, $\mathcal{B}_c$ in case of centralized learning or individual reply buffers, $\left\{\mathcal{B}_n\right\}_{n=1}^N$ in distributed learning.  

To stabilize the learning process, MADDQN utilizes target networks. These networks are copies of the local Q-networks and their weights are periodically updated by polyak averaging, which involves updating the target network weights with a fraction of the local network weights.

 The Q-networks are optimized using mini-batches of experiences sampled from the replay buffer with the goal of learning decentralized policies that maximize the expected cumulative rewards for each agent. For each agent, the Q-network is trained to minimize the mean squared Bellman error, i.e., the squared difference between the estimated Q-value and the target Q-value. The loss function is defined as \cite{feriani2021single}
 \begin{equation}
     \mathcal{L}(\theta) = \frac{1}{N} \sum_{i=1}^{N} (y_i - Q(s_i, a_i; \theta_i))^2
 \end{equation}
where $N$ is the mini-batch size, $s_i$ is the state, $a_i$ is the action, and $Q(s_i, a_i; \theta_i)$ is the estimated Q-value from the local Q-network. The variable, $y_i$ is the target Q-value which is calculated by using the target network to select the best action for the next state and then evaluating it with the local Q-network. The target Q-value is typically defined as 
\begin{equation}
    \label{eq:bellmantarget}
    y_i = r_i + \gamma Q'\left(s', \max_a Q(s', a'; \theta_i);\theta_i'\right)
\end{equation}
where $r_i$ denotes the reward received by agent $i$, $s'$ is the next state, $a'$ is the action selected by the target Q-network, and $Q'$ is the target Q-network.

During training or execution, each agent selects actions using its local Q-network based on the observed state using the well-known $\epsilon$-greedy strategy. The actions can be either random with probability, $\epsilon$, to encourage exploration, or greedy with probability, $1-\epsilon$, to exploit the learned policies. 

\subsubsection{MAPPO}

Multi-agent Proximal Policy Optimization (MAPPO) is a state-of-the-art algorithm designed to address challenges in multi-agent environments. It is an extension of the Proximal Policy Optimization algorithm, originally proposed for single-agent reinforcement learning. PPO uses an actor-critic architecture, where each agent has its own policy and value function, allowing them to interact and learn decentralizedly. The core idea is to update the policy while ensuring deviation from the previous policy remains within a certain range. This is achieved by maximizing a surrogate objective function, which approximates the expected improvement of the policy using the current state-action distribution and the ratio between new and old policies. This balances exploration and exploitation, enabling stable learning in multi-agent environments.

In MAPPO, each agent interacts with the environment by executing actions according to its policy. Trajectories of state-action pairs are collected and stored in a replay buffer for efficient sampling. The collected trajectories are used to estimate the value function for each agent, typically using a separate critic network to predict cumulative rewards from each state. The surrogate objective is computed based on the collected trajectories and the current policy, quantifying the expected improvement and determining the policy update direction.
 Denoting the ratio of the new policy, $\pi_{\theta'}'$ and the old policy, $\pi(\theta)$ as $r(\theta)$, the surrogate objective is defined as \cite{kuba2021trust}
\begin{equation}
    \label{surrogateObjective}
    \mathcal{L}(\theta) = \mathbb{E}\left[\min\left(r(\theta')A, \operatorname{clip}\left(r(\theta',1-\eta,1+\eta\right)A\right)\right],
\end{equation}
where $A$ denotes the advantage function which provides a measure of the advantage of taking a specific action under the current policy compared to the estimated value function. The parameter $\eta$ is a hyperparameter that controls the range of the policy deviation.

The policy is updated by optimizing the surrogate objective function using stochastic gradient descent (SGD) or a similar optimization technique. The objective is to maximize the surrogate objective with respect to the policy parameters, $\theta$. This is typically done by taking multiple gradient steps on the objective function, considering the trajectories collected from the replay buffer.

The update of the parameters, $\theta$ of the policy can be represented by
\begin{align}
    \label{policyupdate}
    \theta' &=\arg\max_{\theta'}\left\{\frac{1}{N}\sum_{m=1}^N\min\left(r(\theta')A_m, C(\theta') A_m\right)\right\},\\
    C(\theta') &=\operatorname{clip}\left(r(\theta',1-\eta,1+\eta\right),
\end{align}
where $N$ denotes the number of trajectories and $A_m$ is the advantage of the $m^\text{th}$ state-action pair.

The value function is updated to improve its estimation accuracy. This is typically done by minimizing the Mean-Squared Error (MSE) between the predicted value function and the actual cumulative rewards. The value function update is performed using SGD or another optimization algorithm, and the target is to minimize the loss function:
\begin{equation}
    \label{valuefunctionupdate}
    \mathcal{L}(\theta_v) = \frac{1}{N}\sum_{i=1}^{N}(V(s_i;\theta_v) - G_i)^2,
\end{equation}
where $\theta_v$ represents the value function parameters, $s_i$ is the $i^\text{th}$ state, $V_i$ is the predicted value function for state $s_i$, and $G_i$ is the actual cumulative reward obtained from the $i^\text{th}$ trajectory.

The steps above are repeated iteratively to improve the policy and value function over multiple iterations. The process continues until a desired level of performance or convergence is achieved.

\subsection{Proposed Methods}
We propose two algorithms combining the MARL techniques, i.e., MADDQN and MAPPO described above with federated learning \cite{collins2022fedavg,mansour2022federated} to solve the resource optimization problem in \eqref{eqPrb}. The methods are referred to as Federated MADDQN (F-MADDQN) and Federated MAPPO (F-MAPPO). The algorithms require the definition of the environment, action space, state space, and reward function. The considered environment is the wireless network with $N$ in-X subnetworks. The other components of the proposed solutions are described below.

\subsubsection{Action Space}
 As stated earlier, we assume that the available frequency band is divided into a set of $K$ equally sized channels. This leads to a $K$-dimensional action space for each subnetwork. Denoting the $k^\text{th}$ channel as $b_k$, the action space for the $n^\text{th}$ subnetwork, $\mathcal{A}^n$ is defined as 
\begin{equation}\label{eq:channel_action_space}
    \mathcal{A}^n = \left\{b_1,\ldots,b_K\right\}\quad \forall n\in\mathcal{N} 
\end{equation}
\subsubsection{Observation Space}
The local observation of subnetwork $n$ at time $t$ is defined as
\begin{equation}
    \mathbf{S}_n(t) = \left[f(\mathbf{s}_n^1(t)),\cdots,f(\mathbf{s}_n^K(t))\right]^T,
\end{equation}
where $\mathbf{s}_n^k(t) = [\mathrm{SIR}_{n1}^k(t), \cdots, \mathrm{SIR}_{nM}^k(t)]\in \mathbb{R}^{M\times 1}$ denotes the vector of SIRs measured at all devices in the $n^\text{th}$ subnetwork on channel $k$ and $f$ is the Observation Reduction Function (ORF) which is applied to the measurements to obtain the state to be used as input to an RL agent. The choice of ORF affects the input to the RL model and hence plays a significant role in the learning process. Considering the minimum rate constraint in \eqref{eqPrb}, it appears reasonable to optimize for the worst case, i.e., allocate channels based on the minimum SIR over each channel. However, the combinatorial nature of the optimization problem makes it difficult to conclude whether this is the optimal choice. We therefore propose to use two categories of ORF defined as
\begin{equation}\label{eq:ORF}
    f(\mathbf{x}) : 
    \begin{cases}
        \mathbf{x} \mapsto \mathbf{x}\in\mathbb{R}^{M\times 1} & \text{Case I:Full state}\\    
        \mathbf{x} \mapsto f_A(\mathbf{x})\in\mathbb{R}^{1} & \text{Case II:Reduced state}
    \end{cases},
\end{equation}
where $f_A$ is the mean, max, or median aggregation functions.

\subsubsection{Reward Signal}
To guide the learning of the agents toward achieving this objective in \eqref{eqPrb}, we define the reward function for the $n$ subnetwork as

\begin{equation}\label{eqReward11}
     R_n = \lambda_1\sum_{m=1}^M r_{n,m} - \lambda_2\sum_{m=1}^M \mathds{1}(r_{n,m})(r_{\mathrm{min}}-r_{n,m})
\end{equation}
 
where $\lambda_i; i=1,2$ are scaling factors that are selected to create a balance between rate maximization and satisfaction of the minimum rate constraint. The function, $\mathds{1}(r)$, is a binary indicator function with a value equal to unity if and only if $r\geq r_{\mathrm{min}}$. 

\subsubsection{Policy}
The main component of any reinforcement learning solution is the policy that defines how state measurements from the environment are mapped into actions. Representation of this state-action mapping and the associated optimization framework determines the policy. In this paper, we propose two methods viz: F-MADDQN and F-MAPPO. As stated earlier F-MADDQN and F-MAPPO combines federated learning with MADDQN and MAPPO, respectively. In both algorithms, the policy is modeled using a DNN.

\subsubsection{Training}
In both F-MADDQN and F-MAPPO, the agents are trained using federated learning (FL) \cite{5_federated} as shown in Figure~\ref{fig1}c. Consider $N$ agents $\left\{\mathcal{F}_i\right\}_{i=1}^N$, each storing their respective data-sets in experience replay buffer, $\mathcal{B}_i$ from which a set of model parameters $\mathbf{\theta}_i$ is learned. Denoting the loss function for the $i^\text{th}$ agent as $\mathcal{L}_i(\mathbf{\theta}_i)$. the common global model loss can be defined as $\mathcal{L}_g(\mathbf{\theta})$ \cite{5_federated}.
\begin{equation}
   \mathcal{L}_g(\mathbf{\theta}) = \sum_{i=1}^N \eta_i \mathcal{L}_i(\mathbf{\theta}),
\end{equation}
where $|\cdot|$ denotes the size of the set and $\eta_i>0$ is the relative impact of each agent. Typically, the term $\eta$ is constrained to $\sum_{i=1}^N \eta_i=1$.  To allocate equal priority to all agents, we use $\eta_i=1/N; \forall i\in\mathcal{N}$. The goal is to find the optimal parameters $w^*$, which minimizes the global loss function \cite{5_federated}.
\begin{equation}\label{eq:fl_optimisation_problem}
    \mathbf{\theta}^* = \arg\min_{\mathbf{\theta}} \mathcal{L}_g(\mathbf{\theta})
\end{equation}
One solution for \eqref{eq:fl_optimisation_problem} would be a gradient-descent approach, known as the federated averaging algorithm. Each agent uses its local data to perform a number of steps in gradient descent on current model parameters $\bar{\mathbf{\theta}}(t)$. This gradient descent step is defined as
\begin{equation}
    \mathbf{\theta}_i(t+1) = \bar{\mathbf{\theta}}(t) - \gamma\nabla \mathcal{L}_i(\bar{\mathbf{\theta}}_i(t))\quad \forall i\in\mathcal{N},
\end{equation}
where $\gamma>0$ is the learning rate, and $\nabla f(\mathbf{\theta})$ for any scalar expression $f(\mathbf{\theta})$ denotes the vector of partial derivatives with respect to the components of the parameters $\mathbf{\theta}$. During federated training, the global model is updated as
\begin{equation}
    \mathbf{\theta}_g(t+1) = \sum_{i=1}^N \dfrac{1}{N} \mathbf{\theta}_i(t+1)
\end{equation}
Once updated, the global weight is sent back to the agents at specified intervals referred to as \emph{aggregation interval}, $\tau_{\mathrm{agg}}$.

\begin{table}
\caption{Simulation parameters.}
    \centering
    \setlength{\tabcolsep}{1pt}
    \renewcommand{\arraystretch}{1}
  \resizebox{0.8\columnwidth}{!}{  \begin{tabular}{|ll|l|}
        \hline
        \textbf{Parameter} && \textbf{Value} \\
        \hline
        Total factory area & $\mathcal{R}$ & $180~\mathrm{m}\times 80~\mathrm{m}$ \\
        \hline
        Clutter type table &  & Sparse \\
        \hline
        Number of subnetworks & $N$ & 20 \\
        \hline
        Timestep & $t$ & 0.005~s \\
        \hline
        Number of episode & & 2000 \\
        \hline
        Number of steps per episode & $T$ & 200 \\
        \hline
        Subnetwork separation distance & $d_{\text{min}}$ & 1~m \\
        \hline
        Subnetwork radius & $d_r$ & 1~m \\
        \hline
        Subnetwork velocity & & 3~m/s \\
        \hline
        Transmit power & $p_n(t)$ & -10~dBm \\
        \hline
        Number of frequency channels & $K$ & 4  \\
        \hline
        Carrier frequency & $f_c$ & 6~GHz \\
        \hline
        Bandwidth per subnetwork & $BW$ & 10~MHz \\
        \hline
        Noise figure & NF & 10~dB \\
        \hline
        Shadowing decorrelation distance & $d_{\delta} $ & 10~m \\
        \hline        
        Max action switch delay & $\tau_{\text{max}}$ & 10 \\
        \hline
    \end{tabular}}
    
    \label{tab:env_parameters}
\end{table}
\begin{figure*}
    \centering
    \includegraphics[width=0.8\textwidth]{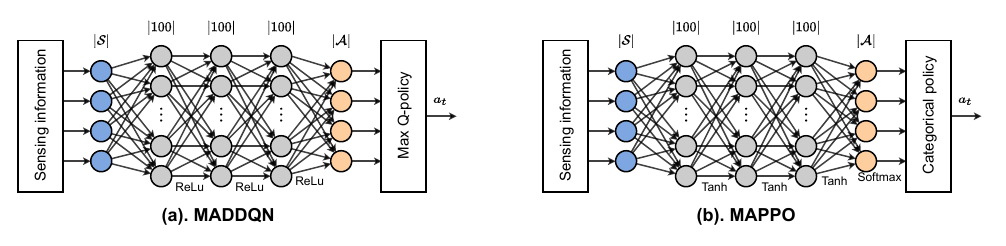}
    \caption{Illustration of the (a) Deep Q-Network (DQN) architecture for MADDQN and (b) Deep PPO architecture for MAPPO which are used for the simulations.}
    \label{fig:architectures}
\end{figure*}
\section{PERFORMANCE EVALUATION}\label{sec:4_PERFORMANCE_EVALUATION}
\subsection{Simulation Settings}
We consider an indoor industrial factory scenario inspired by existing production facilities of manufacturing companies as identified by the industry initiative, 5G alliance for connected industries, and automation \cite{5gacia5GACIAAlliance}. 
Such a layout is implemented as a $180~\mathrm{m}\times 80~\mathrm{m}$ hall containing several separate areas for production, assembly, storage, and human work zones. Multiple in-robot subnetworks can be deployed with the task of transporting materials or tools around the facility.
The alleys separating laboring areas are 5 m wide taking up $\sim 1600$ m$^2$ of the factory area and are outlined as two-lane roads in a right-handed traffic setting.
In the deployment, the in-robot subnetworks are separated with a minimum distance of $d_{\text{min}}=1$ m and move with a speed of 3~m/s. 
The in-robot subnetworks are modeled by a circular coverage area with a radius of $d_r=1$ m, making it possible for robots to pass each other in the alleys. Collisions are avoided by prioritizing robots with the shortest distances to a common intersection, slowing down any other robot that is within minimum separation distance.

Except where otherwise stated, we consider a total bandwidth of $100~\mathrm{MHz}$ which is partitioned into $K = 4$ channels. Transmission within each subnetwork is then performed over a single channel at each time instant. The channel gain is calculated as $g_{n,m} = h_{n,m} \times \sqrt{10^{(PL_{n,m}+X^{\mathrm{SD}}_{n,m})/10}}$, where $h_{n,m}$ denotes the small scale channel gain which is modeled as a temporally correlated Rayleigh random variable, $PL_{n,m}$ denotes the path loss and $X^{\mathrm{SD}}_{n,m}$ is the shadow fading. We simulate the pathloss using the 3GPP Indoor Factory (InF) channel model for the Dense-clutter Low-antenna (InF-DL) and the Sparse-clutter Low-antenna (InF-SL) scenarios. The shadow fading $X^{\mathrm{SD}}_{n,m}$ is generated using the spatially correlated shadowing model used in \cite{adeogun2022multi}. Other simulation parameters are given in Table~\ref{tab:env_parameters}. We consider different values of the aggregation intervals, i.e.,  $T^{\text{Agg}} =[128,256,512,1024]$ for F-MAPPO and F-MADDQN. The minimum required data rate is set as, $r_{\rm{min}} = 11~\mathrm{bps/Hz}$.
\subsection{Benchmarks}
We benchmark the performance of the proposed F-MADDQN and F-MAPPO algorithms with the following methods. \\
1. \textbf{Centralized Graph Coloring (CGC)}: The CGC algorithm utilizes improper coloring to assign colors equivalent to channels for all subnetworks \cite{litC4}.
At each timestep $t$, the pair-wise interference power relationships among subnetworks, $\pmb{I}(t)\in\mathbb{R}^{N\times N}$, are collected, and mapped to a mutual coupling graph $G_t$.
Each vertex corresponds to a subnetwork, and edges are created by connecting each vertex to its $K-1$ nearest neighbors, where the weights of edges are equivalent to the interference power between the connected subnetworks.\\
2. \textbf{Greedy channel selection}: At each switching instant, each subnetwork selects the channel with the highest measured SINR from the previous time step. 
 \\
3. \textbf{Random channel selection}:  At the beginning of every episode, a random channel is allocated to each subnetwork. 

In addition, the federated learning solutions are also compared with previous solutions based on centralized as well as distributed training. We denote the centralized (distributed) MADDQN and MAPPO as C-MADDQN (D-MADDQN) and C-MAPPO (D-MAPPO), respectively.
\begin{figure*} 
    \centering
  \subfloat[MADDQN.]{
        \includegraphics[width=0.42\textwidth] {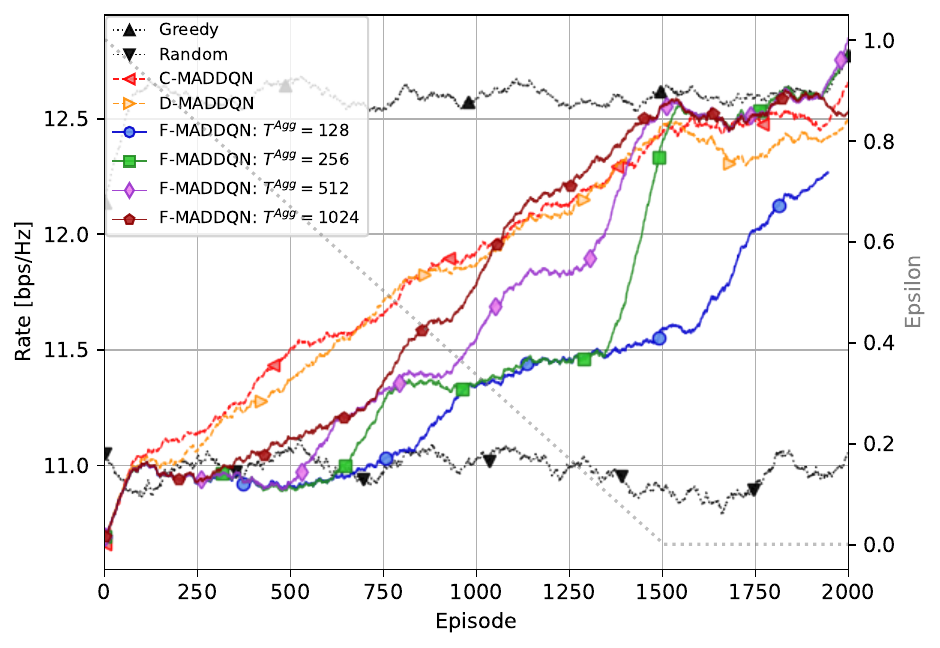}        \label{fig:training_rate_maddqn} } 
          \subfloat[MAPPO.]{
\includegraphics[width=0.42\textwidth] {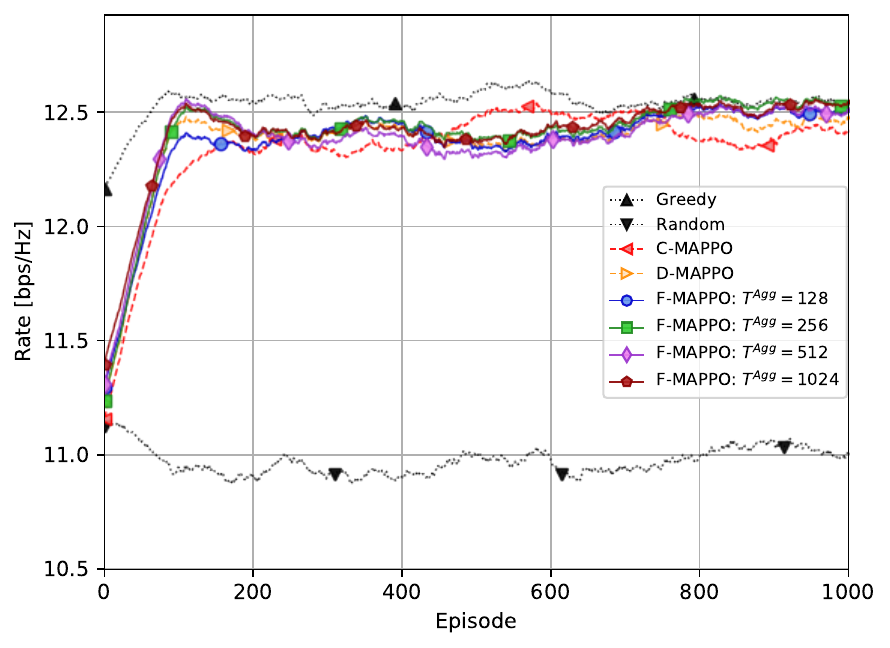}
     \label{fig:training_rate_mappo} }\\
  \caption{Averaged reward versus number of episodes.}
  \label{fig:trainingcurves} \vspace{-15pt}
\end{figure*}
\begin{figure}
    \centering
    \includegraphics[width=0.38\textwidth]{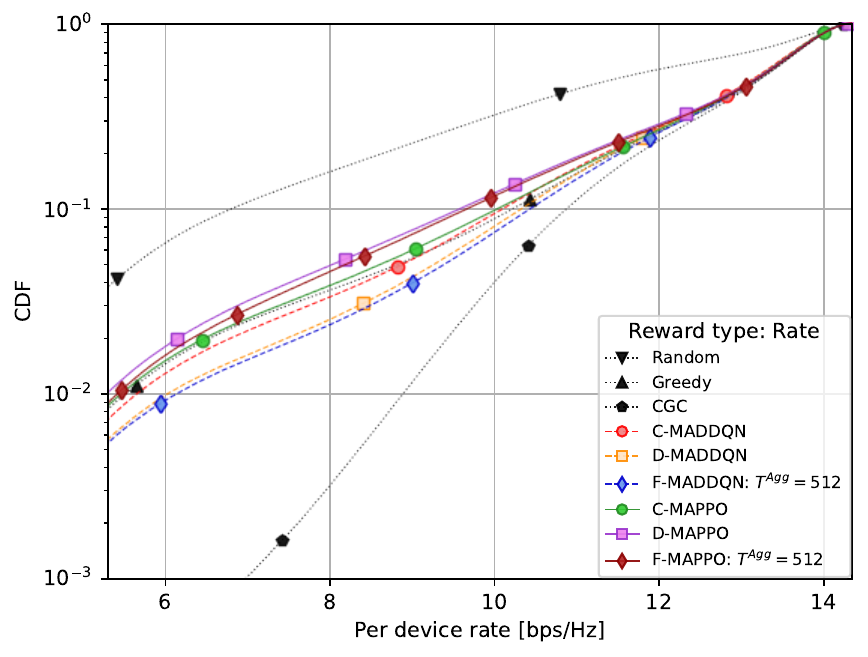}
    \caption{CDF of achieved rate.}
    \label{fig:evaluation}
\end{figure}
\begin{figure}[t]
    \centering
    \includegraphics[width=0.38\textwidth]{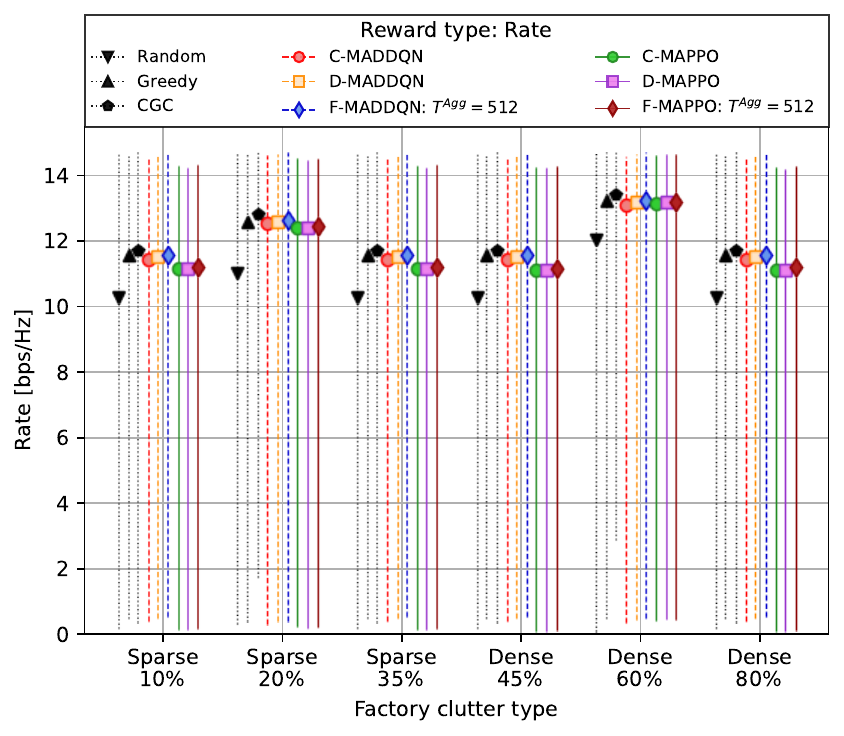}
    \caption{Sensitivity evaluation.}
    \label{fig:sensitivity}\vspace{-10pt}
\end{figure}
\begin{figure}
    \centering
    \includegraphics[width=0.38\textwidth]{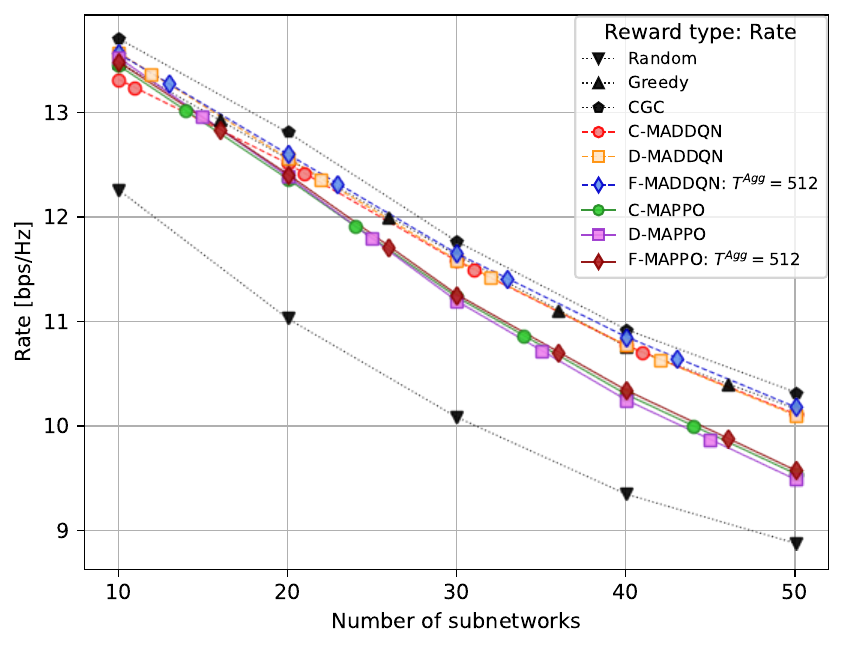}
    \caption{Rate versus number of subnetworks.}
    \label{fig:density}
\end{figure}

\subsection{Training and Convergence}
We consider the network architecture in Figure~\ref{fig:architectures}a and Figure~\ref{fig:architectures}b for F-MADDQN and F-MAPPO, respectively. The networks are trained using the federated learning procedure illustrated in Figure~\ref{fig1}c. Figure~\ref{fig:training_rate_maddqn} shows the averaged reward signal achieved at each training episode by the MADDQN-based methods.  At convergence approximately after 1500 episodes, the MARL methods approach the distributed greedy selection baseline. At convergence, the F-MADDQN methods with greater aggregation intervals achieve a marginally better performance than the centralized and distributed baselines, indicating a potential advantage of the federated learning framework. The figure also shows that too frequent aggregation of DQN weights due to a low value of aggregation interval, $\tau_{\mathrm{agg}}$ can result in a comparatively slower convergence rate. For example, while F-MADDQN with $\tau_{\mathrm{agg}}\geq 256$ converged at $\approx 1500$ episodes, the algorithm failed to converge after $2000$~episodes with $\tau_{\mathrm{agg}}=128$.  
 
We evaluate the convergence performance of F-MAPPO with the in Figure~\ref{fig:training_rate_mappo}.At convergence, achieved approximately after $\approx 750$ episodes, the F-MAPPO achieves similar performance to the benchmark for all values of the aggregation interval, $\tau_{\mathrm{agg}}$. Compared to the methods based on MADDQN, the MAPPO-based algorithms give a reduction of about $50\%$ in the number of episodes required for convergence.
A plausible explanation for this is the effect of the decay rate of the $\epsilon$-greedy parameter of MADDQN and the iterative training over multiple epochs at each training step of MAPPO.

\subsection{Performance Comparison}
The trained models are deployed in each in-robot subnetwork for distributed channel allocation, and the performance is compared with random, greedy, CGC, and centralized and distributed training frameworks. To evaluate the trained models, the aggregation interval is set to $T^{\text{Agg}}=512$. Figure~\ref{fig:evaluation} shows the CDF of the achieved rate per device. The proposed F-MADDQN performs marginally better than D-MADDQN and C-MADDQN below the $30^{\text{th}}$ percentile. Furthermore, the performance of F-MADDQN is also better than that of the distributed greedy scheme.
F-MAPPO similarly achieves marginally better performance than the distributed baseline and approaches the centralized baseline below the $30^{\text{th}}$ percentile. As expected, the centralized graph coloring baseline shows significant performance superiority to all other schemes which are based on distributed execution. This difference is expected since CGC exploits the global information about the scenario, unlike the MADDQN, MAPPO, and greedy schemes which rely solely on local measurements of only the aggregate interference power. It should, however, be noted that the CGC method has significantly higher sensing, signalling, and computational complexity. Such high complexity makes CGC impractical for dense deployments of subnetworks.

\subsection{Sensitivity and Robustness Evaluation}
We evaluate the ability of the proposed methods to generalize to changes in the deployment density and the wireless environment parameters. 
\subsubsection{Robustness to changes in deployment density}
The models trained in the factory with $N=20$ subnetworks are deployed in scenarios with different numbers of subnetworks. We vary the number of subnetworks in the test scenarios between $N=10$ and $N=50$ and evaluate the average per-device rate. 
As shown in Figure~\ref{fig:density}, when the number of subnetworks increases, an overall decrease in performance with similar proportions across all methods is observed. While the methods based on MADDQN appear to be robust to the changes in deployment density for all values of $N$, there seems to be a degradation in the performance of the methods based on MAPPO. This shows that the proposed F-MADDQN is more robust to changes in deployment density than the F-MAPPO. 
\subsubsection{Sensitivity to changes in environment parameters}
To study the ability of the proposed methods to generalize to environments with different parameters than those used for the training, we train the models in a factory environment with sparse clutter with an average clutter element size of $d_{\text{clutter}}=10$~m and a density of $r_{\text{clutter}}=20\%$.
The models are then introduced to environments with different types of clutter. We consider sparse and dense clutter cases with varying clutter density as defined in the 3GPP model for in-factory environment \cite{pathloss_3gpp}. Figure~\ref{fig:sensitivity} shows the minimum, maximum, and average values of the achieved rate per device for the different cases using the proposed schemes and the baseline algorithms. The figure shows that the RL-based methods can maintain their performance in comparison to the centralized and distributed baselines.
Furthermore, similar to the observation from the sensitivity test with varying numbers of subnetworks, the MADDQN-based solutions appear to be slightly more robust to changes in the environment and maintain performance close to the greedy selection baseline.

\section{CONCLUSION}\label{sec:5_CONCLUSION}
We proposed two federated reinforcement learning-based schemes named Federated Multi-Agent Double DQN (F-MADDQN) and Federated Multi-Agent Deep Proximal Policy Optimization (F-MADPPO)  for dynamic channel allocation in 6G in-X subnetworks which overcome the inherent issues of convergence and high signalling overhead with privacy challenges associated with distributed and centralized multi-agent reinforcement learning techniques. Our performance evaluation results using realistic in-factory models defined by 5G-ACIA and 3GPP have shown that the proposed schemes can achieve similar performance to the best-performing baselines and are robust to changes in the deployment density as well as wireless environment conditions. 
\vspace{-1pt}\bibliographystyle{IEEEtran}
\bibliography{ref}

\end{document}